\documentclass[aps,prd,twocolumn,tightenlines,showpacs,amsmath,amssymb,nofootinbib]{revtex4-1}

\usepackage{graphicx}
\usepackage{amsmath,amssymb}
\usepackage{bm}
\usepackage{epsfig}
\usepackage{color}              
\usepackage{comment}
\usepackage{hyperref}
\usepackage{array, float,tabularx}
\usepackage{lipsum}

\usepackage{amsmath}
\usepackage{slashed}

\def\be{\begin{equation}}
\def\ee{\end{equation}}
\def\ba{\begin{eqnarray}}
\def\ea{\end{eqnarray}}
\def\bal{\begin{align}}
\def\eal{\end{align}}
\def\bald{\begin{aligned}}
\def\eald{\end{aligned}}

\begin{document}

\title{Breaking Free with AI: The Deconfinement Transition}

\date{\today}

\author{Christian Ermann}
\email{christian.ermann@tufts.edu}
\affiliation{Department of Computer Science, Tufts University, MA, USA}  
\author{Stephen Baker}
\email{stbaker@usc.edu}
\affiliation{Department of Astronautical Engineering, University of Southern California, Los Angeles, CA, USA}
\author{Mohamed M. Anber}
\email{mohamed.anber@durham.ac.uk}
\affiliation{Centre for Particle Theory, Department of Mathematical Sciences, Durham
University, South Road, Durham DH1 3LE, UK}

\begin{abstract}
Employing supervised machine learning techniques, we investigate the deconfinement phase transition within $4$-dimensional $SU(2)$ Yang-Mills (YM) theory, compactified on a small circle and endowed with center-stabilizing potential. This exploration encompasses scenarios both without and with matter in either the fundamental or adjoint representations. Central to our study is a profound duality relationship, intricately mapping the YM theory onto an XY-spin model with $\mathbb Z_p$-preserving perturbations. The parameter $p$ embodies the essence of the matter representation, with values of $p=1$ and $p=4$ for fundamental and adjoint representations, respectively, while $p=2$ corresponds to pure YM theory. The logistic regression method struggles to produce satisfactory results, particularly in predicting the transition temperature. Contrarily, convolutional neural networks (CNNs) exhibit remarkable prowess, effectively foreseeing critical temperatures in cases where $p=2$ and $p=4$. Furthermore, by harnessing CNNs, we compute critical exponents at the transition, aligning favorably with computations grounded in conventional order parameters. Taking our investigation a step further, we use CNNs to lend meaning to phases within YM theory with fundamental matter. Notably, this theory lacks conventional order parameters. Interestingly, CNNs manage to predict a transition temperature in this context. However, the fragility of this prediction under variations in the boundaries of the training window undermines its utility as a robust order parameter. This outcome underscores the constraints inherent in employing supervised machine learning techniques as innovative substitutes for traditional order parameters.
 \end{abstract}

\maketitle

\section{Introduction}
\label{Introduction}

Confinement and mass generation in 4-dimensional Yang-Mills (YM) theory is an open and one of the most challenging problems in physics \cite{Greensite:2011zz}. The difficulty of this problem is attributed to the fact that confinement happens at the strong coupling scale when perturbation analysis becomes of little use. The technical definition of confinement is that the chromoelectric field lines between two probe quarks are collimated in a flux tube and that the potential between the probe quarks increases linearly with their separation.  When YM theory is put in a heat bath, the flux tube melts down at a critical temperature, and the theory exhibits a phase transition (deconfinement) between the confined and deconfined phases. Again, deconfinement occurs in the strongly coupled regime, and no reliable analytical techniques are available to tackle the problem. Both confinement and deconfinement phenomena, though, can be seen in full-scale lattice simulations of YM theory. One of the important tasks of the simulations is to examine the nature of the transition, e.g., second/first order or a smooth crossover, and determine the universality classes of second-ordered transitions.

Studies of phase transitions are based on the classical Landau-Ginsberg paradigm, which requires two essential ingredients:  (1) the invariance of the system under a global symmetry and (2) an order parameter that transforms nontrivially under the symmetry.  In the case of pure $SU(N)$ YM theory, the global symmetry is a $\mathbb Z_N^{(1)}$ $1$-form center symmetry that acts on non-contractable Wilson's loops \cite{Gaiotto:2014kfa}, but otherwise, it leaves the YM action invariant. Adding dynamical matter to YM theory changes the center of the theory. In general, given a matter with a representation of $N$-ality $n$, the center symmetry is $\mathbb Z_{\scriptsize\mbox{gcd}(N,n)}^{(1)}$.  For example, while adjoint matter, $n=0$, retains the full center symmetry, a theory with fundamental matter, $n=1$, does not have a center. Thus, there is no order parameter, and hence, there is no meaningful distinction between the confined and deconfined phases.  

Recently, machine learning (ML) algorithms have emerged as an alternative technique to studying phase transitions; see \cite{Mehta_2019,hao2023physicsinformed} for reviews. In this method, one simply takes a large number of lattice images both in the deep-ordered and deep-disordered phases.  The algorithm is trained to distinguish between the two phases and then can be used to predict the critical (transition) temperature and the critical exponents.  Interestingly, one does not necessarily have to define the critical region with absolute precision. The presence of an actual transition implies that its characteristics remain unaffected by a reasonable adjustment of the boundaries of the critical window.

ML techniques have also been used to study phase transitions in YM theory; see, e.g.,  \cite{Boyda:2020nfh,Favoni:2020reg,Boyda:2022nmh,Sale:2022qfn,Aarts:2023vsf,Zhou:2023pti,Kashiwa:2021ctc,Sehayek:2022lxf,Spitz:2022tul,Antoku:2023uti}.  In this paper, we apply these techniques to study the deconfinement transition in $4$-dimensional $SU(2)$ YM theory (with or without matter) by mapping the original theory to simple $2$-dimensional XY-spin models with $\mathbb Z_p$-preserving perturbations. This is a true mapping (or, if you wish, a duality) between the original YM theory and the XY-spin models via well-under-control effective field theory techniques. This mapping is rather lengthy and technical, and the methods used for the mapping span more than a decade of several works. See, e.g., \cite{Unsal:2007jx,Simic:2010sv,Anber:2011gn,Anber:2012ig,Anber:2013doa,Anber:2018ohz} for the technical details and \cite{Poppitz:2021cxe,Dunne:2016nmc} for reviews. We refrain from discussing the details in this article at great length, giving only a synopsis of the methods used and referring the interested reader to the literature. It is important to emphasize that the XY-spin models exhibit all the symmetries of the original YM theory (with or without matter) and genuinely inherit the important effective degrees of freedom relevant to the deconfinement transition. With all these important features, the XY-spin models are void of the complexities associated with YM theories, e.g., the topological sectors, the difficulty of simulating fermions, etc. Thus, the mapping provides a playground that enables us to examine the ML techniques in studying the deconfinement transition without dealing with all the complexities associated with simulating the full-fledged YM theories.  Different values of $p$ correspond to YM theories with distinct matter content: $p=2,4,1$ correspond to pure YM theory, YM theory with adjoint fermions, and YM theory with fundamental fermions, respectively. As a reference example, we may remove the $\mathbb Z_p$-preserving perturbations to recover the pure XY-spin model. This model has a continuous $U(1)$ symmetry, algebraic long-range order, and exhibits the famous Kosterlitz-Thouless (KT) phase transition \cite{Kosterlitz:1973xp}. 

We apply supervised ML techniques to evaluate their suitability, compared to conventional methods (i.e., the Landau paradigm and order parameters), in detecting the deconfinement transition. In this endeavor, we examine two techniques: logistic regression and convolutional neural networks (CNN). The first method yields excellent results for detecting the KT transition in the pure XY-spin model but fails to detect transitions in XY-spin models with $\mathbb Z_p$-preserving perturbations. The logistic regression method also fails in detecting the Ising model phase transition, as has been reported in the literature. Thus, one sees from all these examples the failure of this method in detecting transitions in systems with discrete symmetries. On the other hand, we show that CNN is reliable in detecting the phase transitions in systems with discrete symmetries. Simply, we train CNN to distinguish between spin images taken from the ordered and disordered phases, excluding the critical region where the transition might happen. Then, by feeding the CNN with images from the critical region, the transition temperature is determined as the temperature at which an image has a fifty-fifty chance to belong to either the ordered or disordered phases. Thus, we can define a predictive function of temperature, $f(T)$, which serves as an alternative to the traditional order parameter, such that the transition temperature $T_c$ is defined via the condition $f(T_c)$=0.5. We may also use the derivative $df(T)/dT$ as the susceptibility associated with $f(T)$, and we show that this quantity peaks near the transition temperature precisely the same way the susceptibility of the magnetization (the traditional order parameter) peaks. Furthermore, we use finite-size scaling along with the susceptibility of the predictive function to calculate the critical exponents of our systems. For example, we examined that this method, when applied to the XY-spin model with $\mathbb Z_4$-preserving perturbation (which is dual to YM theory with adjoint fermions), yields results consistent (within the accuracy we were able to attain) with the critical exponents calculated from more traditional methods.

Long-established within the literature (e.g.,  \cite{Fradkin:1978dv}), it is recognized that YM theory coupled with fermions in the defining (fundamental) representation lacks a clearly defined order parameter. Consequently, the concept of phase transitions in this theory becomes unfeasible. Our investigation addresses this proposition by harnessing machine learning techniques. Notably, we examine the XY-spin model with $p=1$, drawing an equivalence to YM theory with fundamental fermions. Intriguingly, employing CNN trained on images resembling ordered and disordered phases yields a  transition temperature for the system. However, our findings reveal that this temperature's stability is limited, contingent upon how we define the boundaries of the training data window. This crucially undermines the ascription of genuine transitional behavior to this theory. This contrasts with the situation in XY-spin models with $p=2,4$, wherein the critical temperature remains relatively insensitive to the boundaries of the training data window.

This paper is organized as follows. In Section \ref{Theory and formulation}, we review the duality relation that enables us to map the original YM theory into XY-spin models with perturbations. Given the extended literature and the long calculations needed to verify the duality, we skip the derivation but only provide the essential ingredients/physics behind the duality. In section \ref{Deconfinement via supervised learning}, we apply two methods, namely, the logistic regression and convolution neural networks, to study phase transitions in the XY-spin models. We conclude by giving an outlook and possible extensions of our study in Section \ref{Outlook}.


\section{Theory and formulation}
\label{Theory and formulation}

We consider a 4-dimensional $SU(2)$ Yang-Mills (YM) theory on $\mathbb R^3\times \mathbb S^1_L$, where $\mathbb S^1_L$ is a spacial circle\footnote{It is important to distinguish between the spatial and temporal circles. The distinction will be clear as we spell out the details.}  of circumference $L$. We take $L$ to be much smaller than the inverse strong-coupling scale of the theory $\Lambda$, i.e., $L\Lambda\ll 1$. As we argue below, in this limit, the theory becomes weakly coupled and amenable to semi-classical analysis. Throughout this work, we shall consider three cases: pure YM theory, YM theory with adjoint fermions, and YM theory with fundamental fermions. We also impose periodic boundary conditions on $\mathbb S_L^1$ for both gauge fields and fermions\footnote{\label{distinction} Imposing periodic boundary conditions on gauge fields is the standard choice, corresponding to thermal gauge theory. Notice, at this level, that it really makes no difference whether $\mathbb S_L^1$ is temporal or spacial in pure YM theory. The distinction, however, is manifest once we endow the theory with fermions that satisfy periodic boundary conditions. The latter case corresponds to the twisted partition function: ${ \cal Z}=\mbox{tr}\left[e^{-HL}(-1)^F\right]$, where $(-1)^F$ is the fermion number.}. The three different theories will help us make conclusions about the usefulness of the machine learning paradigm in understanding phase transitions in general and deconfinement phase transitions in particular.   

Pure YM theory on a large compact manifold (large compared to $\Lambda^{-1}$) has a $1$-form center symmetry $\mathbb Z_2^{(1)}$ that acts on non-contractible Wilson's loops\footnote{\label{comactification}The prototype example is to think of the $4$-manifold as a $4$-torus. Since the $4$-torus has non-contractible directions (cycles), one can rigorously define non-contractible Wilson's loops and $1$-form symmetries. When the length of the cycles is much larger than $\Lambda^{-1}$, practically speaking, the manifold approaches $\mathbb R^4$. Therefore, throughout this work, when we speak about $\mathbb R^n$ space, the reader should think of a large torus with a cycle length approaching infinity.}. The latter is the order parameter for the confinement/deconfinement phase transition.  Upon compactifying one of the large directions over a small circle, the $1$-form symmetry bifurcates to  $1$-form $\mathbb Z_2^{(1)}$ and  $0$-form $\mathbb Z_2^{(0)}$ symmetries. The $1$-form symmetry acts on spacial Wilson's loops in $\mathbb R^2$ (we imagine compactifying $\mathbb R^{3}$ on a large torus, see Footnote \ref{comactification}), while the  $0$-form symmetry $\mathbb Z_2^{(0)}$ acts on the dimensionally-reduced Polyakov's loops\footnote{The Polyakov's loop wrapping $\mathbb S_L^1$ is given by $\mbox{tr}_\Box \left[e^{i \oint_{\mathbb S_L^1} A^{(1)}}\right]$. When $\mathbb S_L^1$ is small, and the fluctuations of $A^{(1)}$ along $\mathbb S_L^1$ are not sizable, then we may choose $A^{(1)}$ to be in the $\tau^3$ ($\tau^3$ being the third Pauli matrix) color direction and write $\Phi=\oint_{\mathbb S_L^1} A^{(1)}$. Thus, the $1$-form field $A^{(1)}$ along $\mathbb S^1_L$ is dimensionally reduced to a $0$-form field, and the component of the $1$-form symmetry along the same direction becomes a $0$-form symmetry.} that wraps $\mathbb S_L^1$. 

Like pure YM theory, a theory with adjoint fermions will also admit $1$-form center symmetry $\mathbb Z_2^{(1)}$. However, a theory with fundamentals does not admit a center symmetry; one cannot rigorously define the notion of confinement in a theory with fundamentals. One of the tasks of this paper is to examine whether machine learning can provide an alternative or generalized notion of confinement that may be used in this class of theories. 

Since the circle is much smaller than all other length scales, we may try to write down a $3$-dimensional effective field theory by integrating out a tower of heavy Kaluza-Klein excitations along $\mathbb S^1_L$. In the case of pure YM, this results in a thermal field theory with destabilized $0$-form center symmetry; this is the celebrated deconfined phase\footnote{The deconfinement of pure YM happens at a temperature $T=L^{-1}\sim \Lambda$. Above this temperature, the temporal Wilson's loops obey the perimeter rather than the area law.} of pure YM. It is needless to say that we are not interested in this theory since it is far from being weakly coupled. To restore the center symmetry and force the theory into the weak coupling regime, we need to add by hand a double-trace deformation. Let $\Omega= \mbox{tr}_\Box\left[e^{i \Phi}\right]\equiv\mbox{tr}_\Box \left[e^{i \oint_{\mathbb S_L^1} A^{(1)}}\right]$ be the Polyakov loop wrapping $\mathbb S^1_L$, then adding the term $\sum_{n=1}a_n|\Omega^n|^2$, for positive and large enough values of $a_n$, will restore the $0$-form center symmetry. Effectively, adding the double-trace deformation ensures that the total effective potential $V(\Phi)$ is minimized at the center of the Weyl chamber. Therefore, all the W-bosons are massive\footnote{\label{weyl chamber}Without the double-trace deformation, the potential is minimized at the boundary of the Weyl chamber. This has the effect of keeping some gauge modes massless, and thus, the theory stays in its strongly-coupled regime.}. We call this class of theories deformed YM (dYM)

The above arguments apply even if we endow the theory with fundamental fermions; we still need to add a double-trace deformation to stabilize the center.  We denote the theory with fundamentals and deformation by dYM(F). The situation, however, is different for adjoint fermions. If the latter obey periodic boundary conditions on $\mathbb S^1_L$, integrating them out will generate a center-stabilizing potential\footnote{Here we assume that the number of the left-handed Weyl adjoint flavors $n_f>1$. The case $n_f=1$ is the pure supersymmetric YM theory, and we do not consider it here.}. This class of theories is known as QCD(adj).

Whether we add adjoints or double-trace deformations, in both cases, the theory abelianizes. This happens because the adjoint field $\Phi$ breaks $SU(2)$ down to $U(1)$. The $3$-dimensional $U(1)$ theory can be dualized and described by a compact scalar (dual photon) $\sigma$ via the duality relation $F^3_{\mu\nu}=\frac{g^2}{4\pi L}\epsilon_{\mu\nu\alpha}\partial_\alpha\sigma$. The superscript $3$ denotes the color direction, $g$ is the $4$-dimensional coupling constant, and the Greek letters run over $0,1,2$, keeping in mind that we are using a Euclidean description. The photon kinetic energy term is
\begin{eqnarray}
{\cal L}_{U(1)}=\frac{g^2}{16\pi^2 L}\left(\partial_\mu \sigma\right)^2\,,
\end{eqnarray}
and the compact scalar has a period\footnote{This period corresponds to the normalization $\mbox{tr}_\Box \left[t^a t^b\right]=\delta_{ab}$, where $t^a$ are the Lie-algebra generators. In this normalization, the length square of the simple root is $2$.} $\frac{2\pi}{\sqrt{2}}$. 

The story continues, thanks to the existence of monopole and/or composite instantons. These are nonperturbative objects that extremize the Euclidean path integral, and their existence is guaranteed because of the nontrivial second homotopy class $\Pi_2 (SU(2)/U(1))\in \mathbb Z$. The dominating objects in both Pure YM theory and YM theory with fundamentals are monopole instantons. They carry magnetic charge $\pm \sqrt{2}$ under $U(1)$ and have action $S_m=\frac{4\pi^2}{g^2}$. The 't Hooft vertex of the monopole operators, modulo a prefactor, is $e^{\pm i \sqrt{2}\sigma}e^{-S_m}$, where $e^{-S_m}$ is the monopole fugacity.  The dominating objects in YM theory with adjoints are the bions. These are molecules composed of two monopole instantons; they carry twice the magnetic charge $Q_b=\pm 2\sqrt{2}$ and have twice the actions $S_b=\frac{8\pi^2}{g^2}$, i.e., their 't Hooft vortex is $e^{\pm i 2\sqrt{2}\sigma}e^{-S_b}$. The proliferation of monopoles or bions  can be incorporated in the path integral, which results in the $3$-dimensional effective actions:
\begin{eqnarray}
\nonumber
{\cal L}_{\scriptsize\mbox{dYM, dYM(F)}}&=&\frac{g^2}{16\pi^2 L}\left[(\partial_\mu \sigma)^2+e^{-\frac{4\pi^2}{g^2}} \cos\left(\sqrt{2}\sigma\right)\right]\,, \\
\nonumber
{\cal L}_{\scriptsize\mbox{QCD(adj)}}&=&\frac{g^2}{16\pi^2 L}\left[(\partial_\mu \sigma)^2+e^{-\frac{8\pi^2}{g^2}} \cos\left(2\sqrt{2}\sigma\right)\right]\,.\\
\label{3d effective Lag}
\end{eqnarray}
These Lagrangians show that the proliferation of the magnetic charge generates a mass gap. Further studies of these models show in a remarkable way how confinement happens in YM theories on $\mathbb R^3\times \mathbb S_L^1$. 

It is important to emphasize, again, that the Lagrangians in (\ref{3d effective Lag}) describe an effective field theory at zero temperature; $\mathbb S^1_L$ is by no means a thermal circle. Yet, one wonders about the behavior of this theory as we put it in a heat bath at temperature $T$. This amounts to identifying the effective degrees of freedom as we compactify the temporal direction over a circle of circumference $\beta=\frac{1}{T}$. Thus, now the theory lives on $\mathbb R^2\times \mathbb S_L^1\times \mathbb S_\beta^1$. It was realized in a series of works \cite{Unsal:2007jx,Simic:2010sv,Anber:2011gn,Anber:2012ig,Anber:2013doa,Anber:2018ohz}  that the W-bosons\footnote{The W-bosons appear due to the Higgsing of $SU(2)$ at the center of the Weyl chamber, see Footnote \ref{weyl chamber}.}  play an important role at temperatures near the deconfinement transition. These particles are electrically charged, with electric charge $ Q_W=\pm \sqrt{2}$ under $U(1)$, and very heavy, of mass $M_W=\frac{\pi}{L}$, and do not participate in the $3$-dimensional Lagrangian \ref{3d effective Lag} but become important near the transition. The idea is that these bosons come with a Boltzmann suppression factor $e^{-\frac{M_W}{T}}$, which, near the transition, is comparable to the monopole or bion fugacities. Therefore, the problem reduces to studying a $2$-dimensional electric-magnetic Coulomb gas of W-bosons and magnetic charges (either monopoles or bions). The vacuum is dominated by magnetic charges at low temperatures (magnetically disordered phase) and by electric charges at high temperatures (electrically disordered phase). The competition between the electric and magnetic charges is ultimately responsible for the phase transition. In addition to the W-bosons, the first excited Kaluza-Klein fundamental and adjoint fermions contribute to the Coulomb gas. The mass and charges of the fundamentals are $M_F=\frac{\pi}{2L}$, $Q_F=\pm \frac{1}{\sqrt 2}$, while those of the adjoint fermions are equal to the corresponding values of the W-boson:  $M_{adj}=\frac{\pi}{L}$, $Q_{adj}=\pm \sqrt 2$.

The electric-magnetic Coulomb gas can be mapped to a $2$-dimensional XY-spin model with $\mathbb Z_p$-preserving perturbations. The partition function of this system is
\begin{eqnarray}
\nonumber
&&{\cal Z}\left[y,p,n=\pm1\right]=\left[\prod_I\int_{0}^{2\pi} d\theta_I \right]e^{-H\tilde T}\,,\\
&&H=-\sum_{\langle I,J \rangle} \cos \left(\theta_I-\theta_J\right)-y \sum_{I}\cos(p\,\theta_I)\,,
\label{main Z}
\end{eqnarray}
and $\tilde T$ is a dimensionless temperature. We may also define a new dimensionless temperature $T\equiv 1/ \tilde T$, warning the reader that the newly defined $T$ is not the physical temperature of the system, but both can be related. Moving forward, the symbol $T$ will denote the dimensionless temperature, and we will explicitly indicate instances when the physical temperature is being employed. The first term in the Hamiltonian $H$ is the kinetic energy. The angular variable $\theta_I$ is the spin, which is localized at site $I$ and takes values $0\leq \theta_I<2\pi$. The sum in the first term is restricted to the nearest-neighbor spins. Notice that the kinetic energy is invariant under the continuous shift symmetry (related to the U(1) symmetry) $\theta_I\rightarrow\theta_I+C$, $C \in [0,2\pi)$. The compactness of the angular variable allows the system to have vortices $\oint d\theta=2\pi n$, $n \in \mathbb Z$. The fugacity of the vortices is implicit and practically controlled by the lattice spacing, i.e., the UV cutoff. Yet,  only the lower-order vortices with winding number $n=\pm 1$ will dominate. The second term in $H$ is the perturbation, which breaks the continuous shift symmetry down to $\mathbb Z_p$. The coefficient $y$ controls the strength of perturbations. 

The reader will notice that the Hamiltonian in Eq. (\ref{main Z}) resembles the system in Eq. (\ref{3d effective Lag}). A dimensional reduction of the latter  
 from $3$ to $2$ dimensions reproduces an almost identical form of Eq. (\ref{main Z}) (after discretization and rescaling of variables). Although this is superficially true, the exact connection between Eqs. (\ref{main Z}) and (\ref{3d effective Lag}) is more involved. For example, the naive dimensional reduction of  QCD(adj) Lagrangian allows for lowest-order vortices to dominate the system; the vortices, in this case, are the fundamental fermions. Obviously, these excitations do not exist in QCD(adj), and thus, the naive dimensional reduction of Eq.  (\ref{3d effective Lag}) does not lead to the desired physical properties. One overcomes this problem by using a T-duality, which maps the electric and magnetic charges to each other. The interested reader is referred to \cite{Anber:2018ohz} for the details of this duality. Here, it suffices to say that Eq. (\ref{main Z}) is the T-dual of  Eq. (\ref{3d effective Lag}); this is true for dYM, dYM(F), and QCD(adj). Now, we mention how the different values of $p$ capture the physics of the $3$ distinct theories we have. 
 \begin{enumerate}
 
\item $p=1$. This case corresponds to dYM(F). The perturbation term $y\sum_{I}\cos\theta_I$ accounts for the fundamental quarks with fugacity $y\tilde T$. In principle, one should also add the term $\sum_{I}\cos 2\theta_I$ to account for the W-bosons. The fugacity of the latter, however, is exponentially suppressed compared to the fundamental fermions\footnote{The fugacity of the fundamentals is $\sim e^{-\frac{M_F}{T}}=e^{-\frac{\pi}{2LT}}$, while the W-bosons fugacity is $\sim e^{-\frac{M_W}{T}}=e^{-\frac{\pi}{LT}}$ (here, $T$ is the physical temperature). We easily see that the latter is exponentially suppressed compared to the former.} and can be ignored. The unit winding vortices $n=\pm 1$ are the magnetic monopoles. This model does not exhibit any symmetry since the shift symmetry $\theta_I\rightarrow \theta_I+C$ is broken by the perturbations to nothing. This is exactly what we expect in a theory with fundamental charges since the theory does not admit a center symmetry. 

\item $p=2$. This corresponds to dYM, where the perturbation term corresponds to the W-bosons. Again, The unit winding vortices are the magnetic monopoles. The theory exhibits the $\mathbb Z_2$ symmetry: $\theta_I\rightarrow \theta_I+\pi$. This is the dimensionally-reduced $0$-form part of the $\mathbb Z_2^{(1)}$ symmetry, the order parameter of the confinement/deconfinement transition. The latter acts on Polyakov's loop that wraps $\mathbb S_\beta^1$.

\item $p=4$. This corresponds to QCD(adj). The perturbation term $y\sum_{I}\cos 2\theta_I$ accounts for both the W-bosons and the first excited Kaluza-Klein mode of adjoint fermions (both have the same mass and hence, the fugacity). The unit winding vortices $n=\pm 1$ are the magnetic bions (one needs to study the T-duality that acts on the electric-magnetic charges to see why this is the case compared to dYM and dYM(F)). QCD(adj) enjoys two symmetries: the $0$-form discrete chiral symmetry $\mathbb Z_2^{d\chi}$ that acts on the adjoint fermions and the $1$-form center symmetry $\mathbb Z_2^{(1)}$ that acts on the Polyakov's loop that wraps $\mathbb S_\beta^1$. As we map the theory into the XY-spin model, $\mathbb Z_2^{d\chi}$ chiral and $\mathbb Z_2$ center are enhanced to $\mathbb Z_4$. 

\end{enumerate}

In all cases, exciting a vortex costs energy ${\cal O}(\tilde T)$ (in dimensionless units), which means that vortices (the magnetic charges) are suppressed at high temperatures. On the other hand, the fugacity of the electric charges $\sim y\tilde T$ increases with temperature $\tilde T$, and therefore, they dominate at high temperatures $\tilde T$. This is exactly the expected behavior in all theories discussed in this work. 

In addition to the above cases, it will also be instructive to include the case of the pure XY-spin model, setting $y=0$. This choice does not correspond to YM theory since it does not account for the electric charges. Pure XY-spin model, however, enjoys an exact continuous shift symmetry, and according to the Mermin-Wagner-Coleman theory, it does not have a true long-range order. Yet, it exhibits the celebrated  Kosterlitz-Thouless (KT) phase transition. 

The traditional method in studying phase transitions is to examine the expectation value of the order parameter, which in this case is given by $\langle \sum_{I}e^{i\theta_I}\rangle$, and its higher moments, e.g., the susceptibility. This method will serve as the basis of comparison with the machine learning techniques that we study in the next section. 

\section{Deconfinement via supervised learning}
\label{Deconfinement via supervised learning}

In this section, we apply the methods of supervised machine learning to the XY-spin models with $\mathbb Z_p$ perturbations, taking $p=1,2,4$. We are also interested in studying the pure XY-spin model\footnote{The proliferation of vortices in this model was used in \cite{PhysRevB.97.045207} to train a neural network to distinguish between the ordered and disordered phases.}. Our investigation aims to answer the following questions. 
\begin{enumerate}

\item We test the suitability of two supervised machine learning techniques to identify phase transitions in our systems: logistic regression and convolution neural network (CNN). In particular, we examine the accuracy of both techniques in determining the critical temperature and its dependence on the optimization parameters.


\item We use the predictability function, which predicts the probability of whether we are in the ordered/disorder phase, as an alternative to the conventional order parameter to calculate the critical exponents of the $p=4$ system.

\item Of particular importance is the $p=1$ case, which corresponds to dYM(F). It is well known that this theory does not have an order parameter, and thus, it can only exhibit a smooth cross-over instead of a sharp transition.  It is interesting to examine whether machine learning techniques can identify a hidden ``order parameter" beyond the classical Landau-Ginsberg paradigm that can distinguish between different phases. 

\end{enumerate}

To train and test our machine-learning algorithms, we generate a set of data points using Monte Carlo techniques and the Metropolis algorithm. We consider our models on a $N\times N$ lattice\footnote{The reader should not confuse the lattice size $N$ with $N$ of the gauge group $SU(N)$.} and generate $10^4$ states at every temperature $T$. The states are recorded in every Monte Carlo sweep and randomized to avoid fitting spurious correlations. 

Both logistic regression and CNN were used to predict the critical temperatures of the simulated systems. In both cases, the classifiers were trained on data from low and high-temperature regions, far away from the critical temperature. After being trained, the models were then used to predict the critical temperature and critical exponents, where the system crosses from the low-temperature phase to the high-temperature phase.

\subsection{Logistic Regression}

Logistic regression is a technique used in classification tasks; for a given state of the XY-spin model, with or without perturbations, we would like to identify the phase, i.e., ordered/disordered. Consider the data set $\{y_i,\bm x_i\}$ with a binary label $y_i\in \{0,1\}$ for the disordered/ordered phase, respectively, and $\bm x_i$ is the flattened 2-dimensional $N\times N$ arrays of the spin-state into a 1-dimensional array. We soften the classifier by considering the sigmoid\footnote{We warn the reader not to confuse the sigmoid function with the dual photon.} function $\sigma(x)=1/(1+\exp(-x))$, which yields a continuous range $0\leq\sigma\leq1$ instead of the binary $\{0,1\}$. Values of $\sigma \in (0,\frac{1}{2})$ are considered to be in the disordered phase, and values of  $\sigma \in (\frac{1}{2},1)$ are in the ordered phase.

\begin{figure}[t]
\begin{center}
\includegraphics[width=87mm]{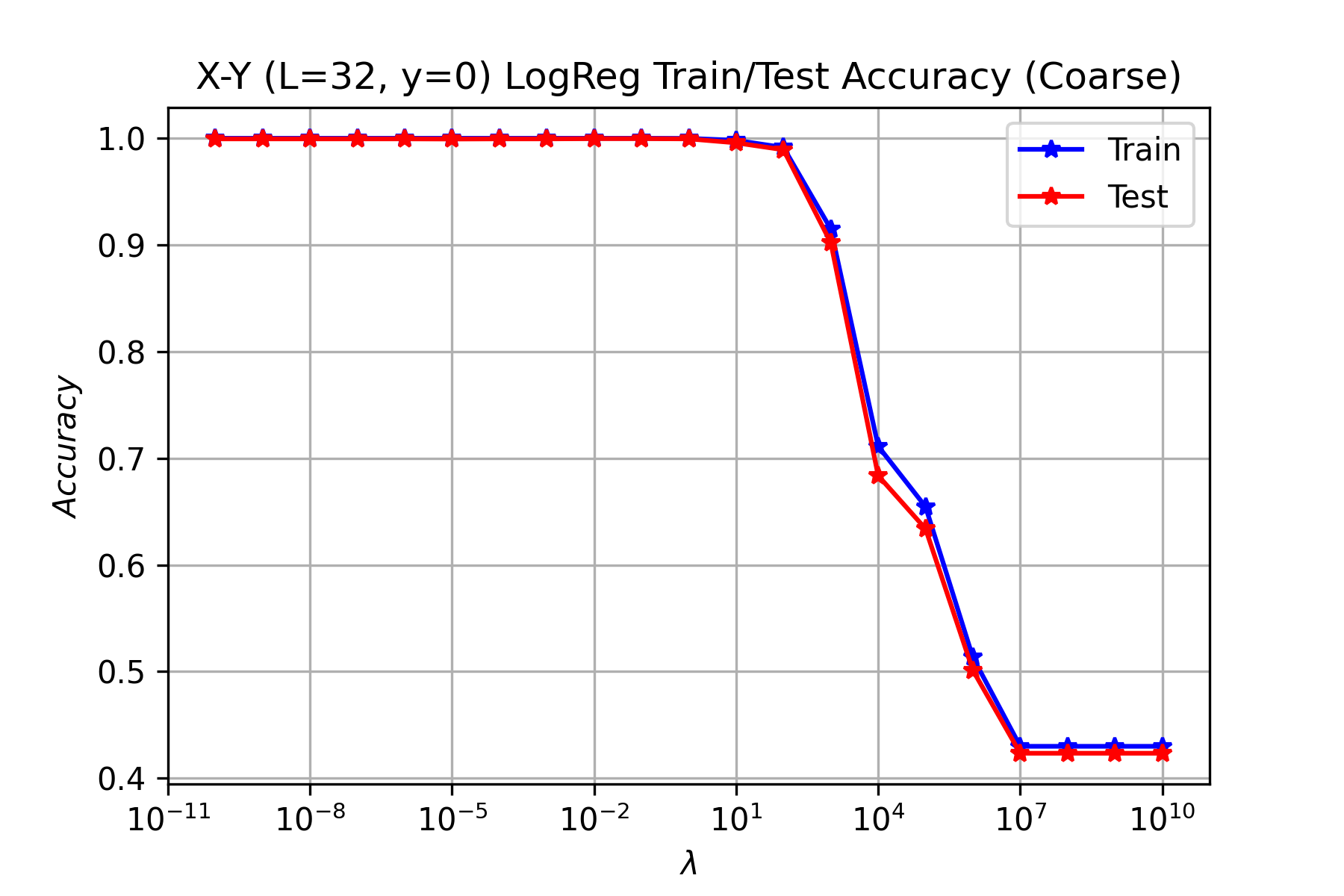}
\includegraphics[width=87mm]{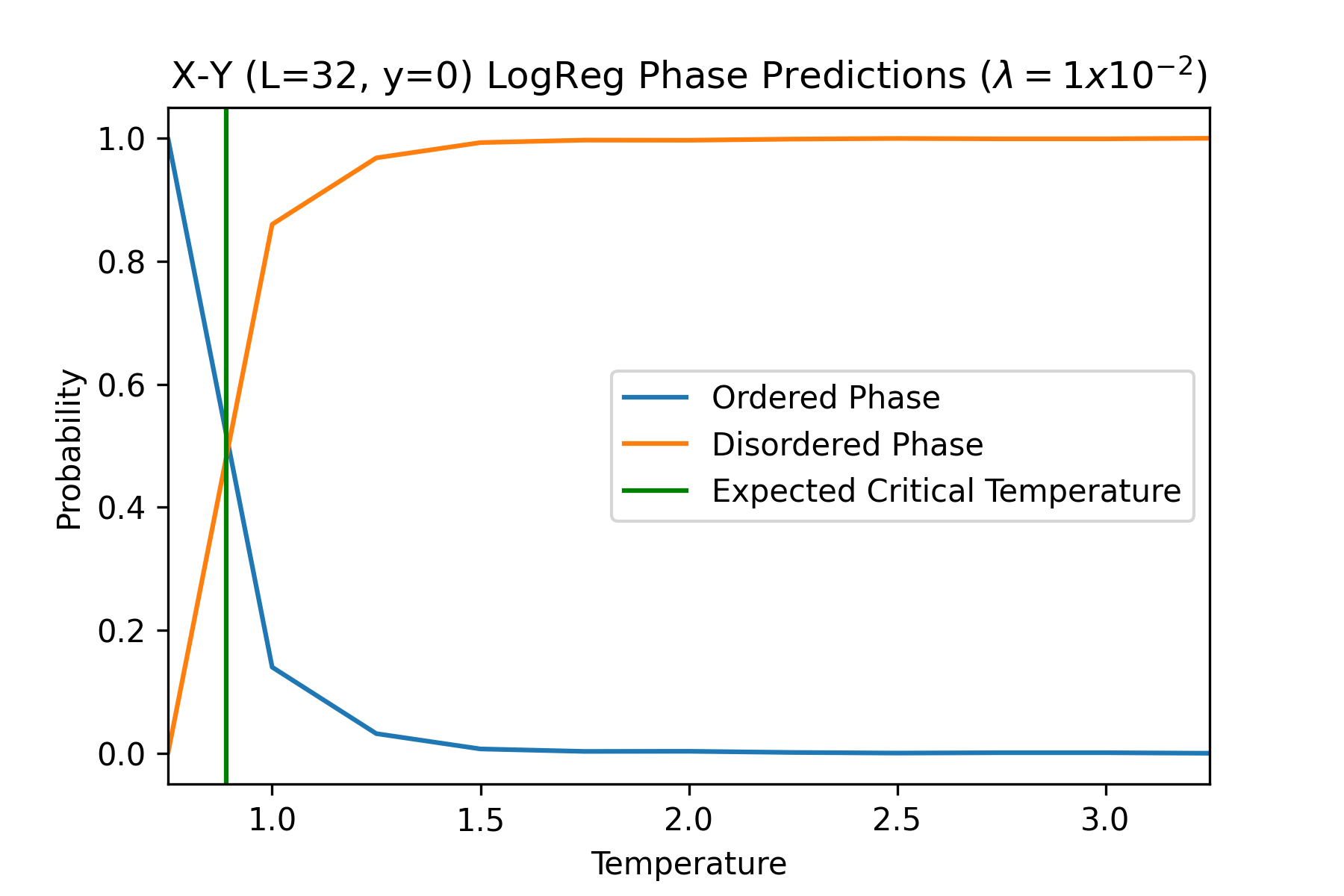}
\caption{Top panel: The accuracy and predictivity of the logistic regression algorithm applied to the pure XY-spin model. We train the algorithm on $n=8\times 10^4$ Monte Carlo configurations in the subcritical and supercritical regions, and then we use the learned weights to predict the transition temperature. Bottom panel: We plot the probabilities versus the dimensionless temperature $T$ that the system is either in the ordered or disordered phase. The intersection between the two probabilities (the $50\%$ chance that the system is in either phase) gives the transition temperature $T_c\cong0.9$, consistent with the literature value of $T_{KT}=0.89$.}
\label{accuracy pure XY}
\end{center}
\end{figure}

According to the Bayesian methods,  the likelihood of observing the data set $\{y_i,\bm x_i\}$ is given by the probability function:
\begin{eqnarray}
\nonumber
P(\{y_i,\bm x_i\}|\bm w)=\prod_{i=1}^n \left[\sigma\left(\bm x_i\cdot \bm w\right)\right]^{y_i}\left[1-\sigma \left(\bm x_i\cdot \bm w\right) \right]^{1-y_i}\,,\\
\label{the full probability}
\end{eqnarray}
for some weights $\bm w$. Then, one readily defines the  cost (error) function (also known as  cross-entropy):
\begin{eqnarray}
\nonumber
C(\bm w)&=&\sum_{i=1}^{n}-y_i\log \sigma\left(\bm x_i\cdot \bm w\right)\\
&&-(1-y_i)\log\left[1-\sigma\left(\bm x_i\cdot \bm w\right)\right]\,.
\end{eqnarray}
 The weights are learned (determined) by minimizing the cost function\footnote{The cross-entropy is a convex function of the wights $\bm w$, and thus, a local minimizer is a global one.}. However, for higher-dimensional data, i.e., $n \gtrsim N^2$, the model might not learn well or overfit\footnote{This can be inferred from calculating the model's accuracy, as we shall see soon.}. To overcome this difficulty, we use the $L^2$ regularization:
\begin{eqnarray}
\hat {\bm w}(\lambda)=\underset{\bm w \in \mathbb R^{N^2}}{\mbox{arg min}}\left(C(\bm w)+\lambda ||\bm w||_2^2 \right)\,,
\end{eqnarray}
with a regularization parameter $\lambda\geq 0$.   The computations are carried out using the Scikit-learn library for machine learning in Python. The minimization procedure is performed with the liblinear routine\footnote{This routine is based on the coordinate descent method, an optimization algorithm that successively minimizes along one coordinate direction at a time until it finds the minimum of a function.}.

We use a lattice size $32\times 32$ and take the temperature $T$ to range from $0.25$ to $4.0$ with a step size of $0.25$, for a total of $16$  temperatures and  $160,000$ configurations. The low-$T$  region is taken in the interval $[0.25, 0.75]$, the high-$T$ region is  $[3.25, 4.00]$, while the critical region is taken in the interval $(0.75, 3.25)$. We divide the data outside the critical region into training and test data. The training and test data sets were formed as a randomly-shuffled combination of the low-T and high-T data sets. $80\%$ of this shuffled combination ($n=8\times 10^4$ configurations) was used for training, and $20\%$ was used for testing.

 We train the logistic regression model on training data of pure XY-spin model and XY-spin model with $\mathbb Z_p$-preserving perturbations, $p=2,4$, to learn the values of $\bm w$ of each model. For a given data point (drawn from the test data) $\bm x_i$, the classifier $\sigma(\bm x_i\cdot \bm w)$ returns the probability, given from Eq. (\ref{the full probability}), of being in the ordered or disordered phase. Then, we define the accuracy of the classifier as the percentage of the correctly identified data.

In FIG.\ref{accuracy pure XY}, we show the accuracy of the classifier for both the test and training data for the pure XY-spin model.  Extremely good accuracy is achieved for a big range of $\lambda$. Then, we fix $\lambda=10^{-5}$ and calculate the predictivity, i.e., the average probability of whether the given set of configurations in the critical region is in the ordered or disordered phase, as a function of temperature. Next, we plot the probabilities that the system is either in the ordered or disordered phase. A $50\%$ chance that the system is in either phase predicts the transition temperature.  The predicted critical temperature is consistent with the literature value of $T_{KT}=0.89$.

\begin{figure}[t]
\begin{center}
\includegraphics[width=87mm]{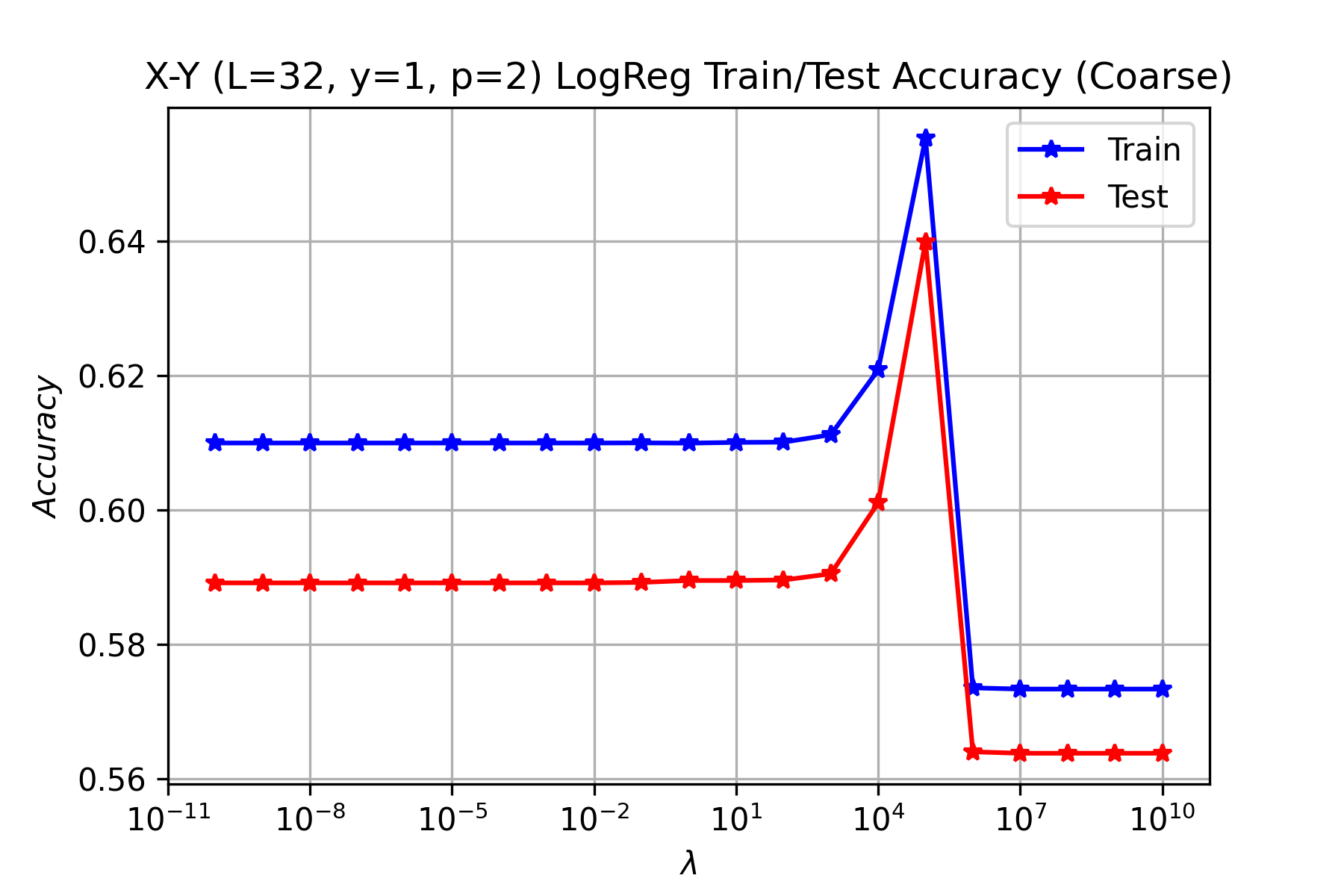}
\includegraphics[width=87mm]{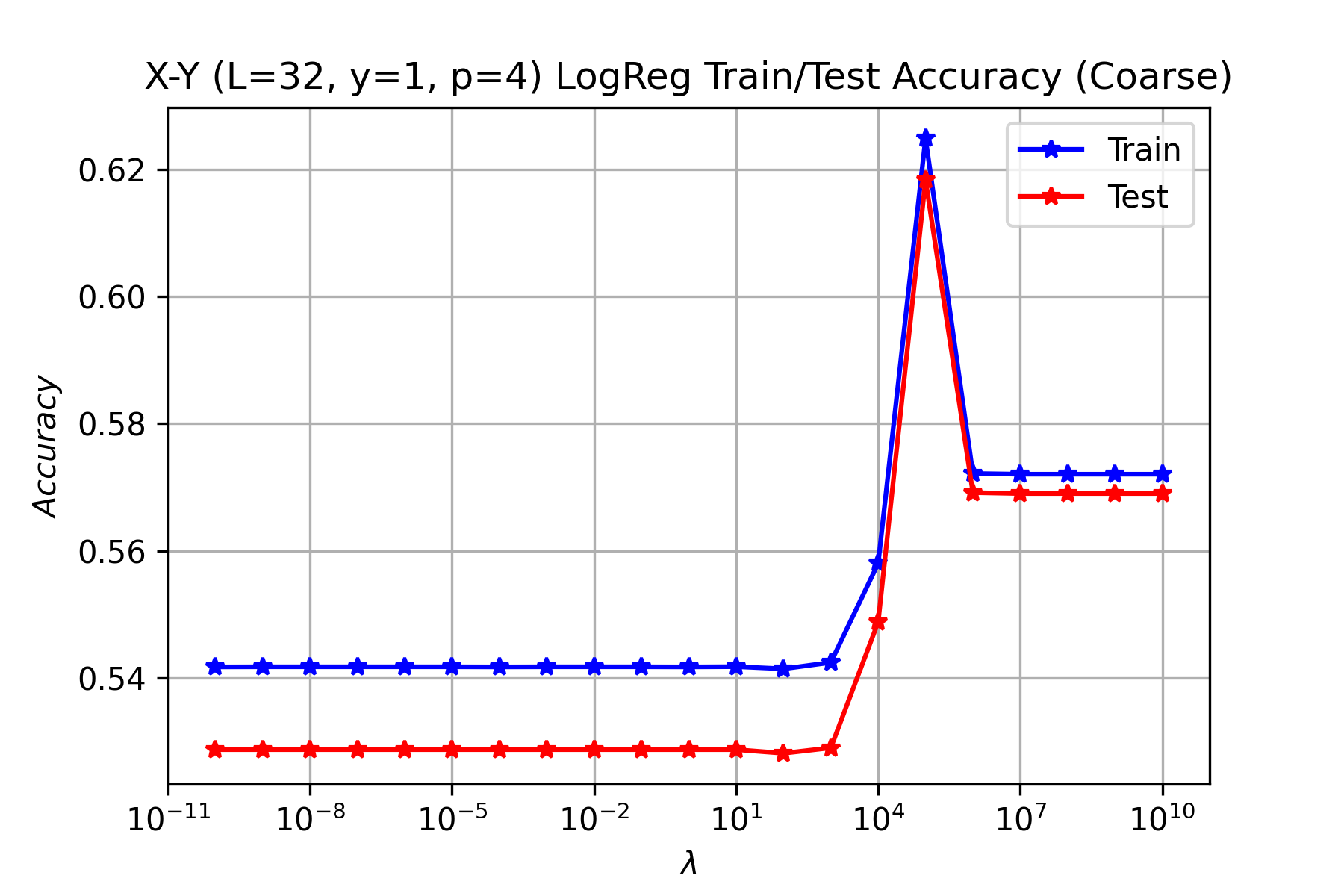}
\caption{The accuracy of the logistic regression algorithm applied to the pure XY-spin model with $\mathbb Z_2$- (top panel) and $\mathbb Z_4$- (bottom panel) preserving perturbations. We train the algorithm on $n=8\times 10^4$ Monte Carlo configurations in the subcritical and supercritical regions. The maximum attained accuracy is below $70\%$ at $\lambda\cong 10^5$, much lower than the accuracy we found for the pure XY-spin model. Such low accuracy hinders the ability of the algorithm to find the transition point.}
\label{accuracy p 2 4}
\end{center}
\end{figure}

\begin{figure*}[ht]
\leftline{
\includegraphics[width=87mm]{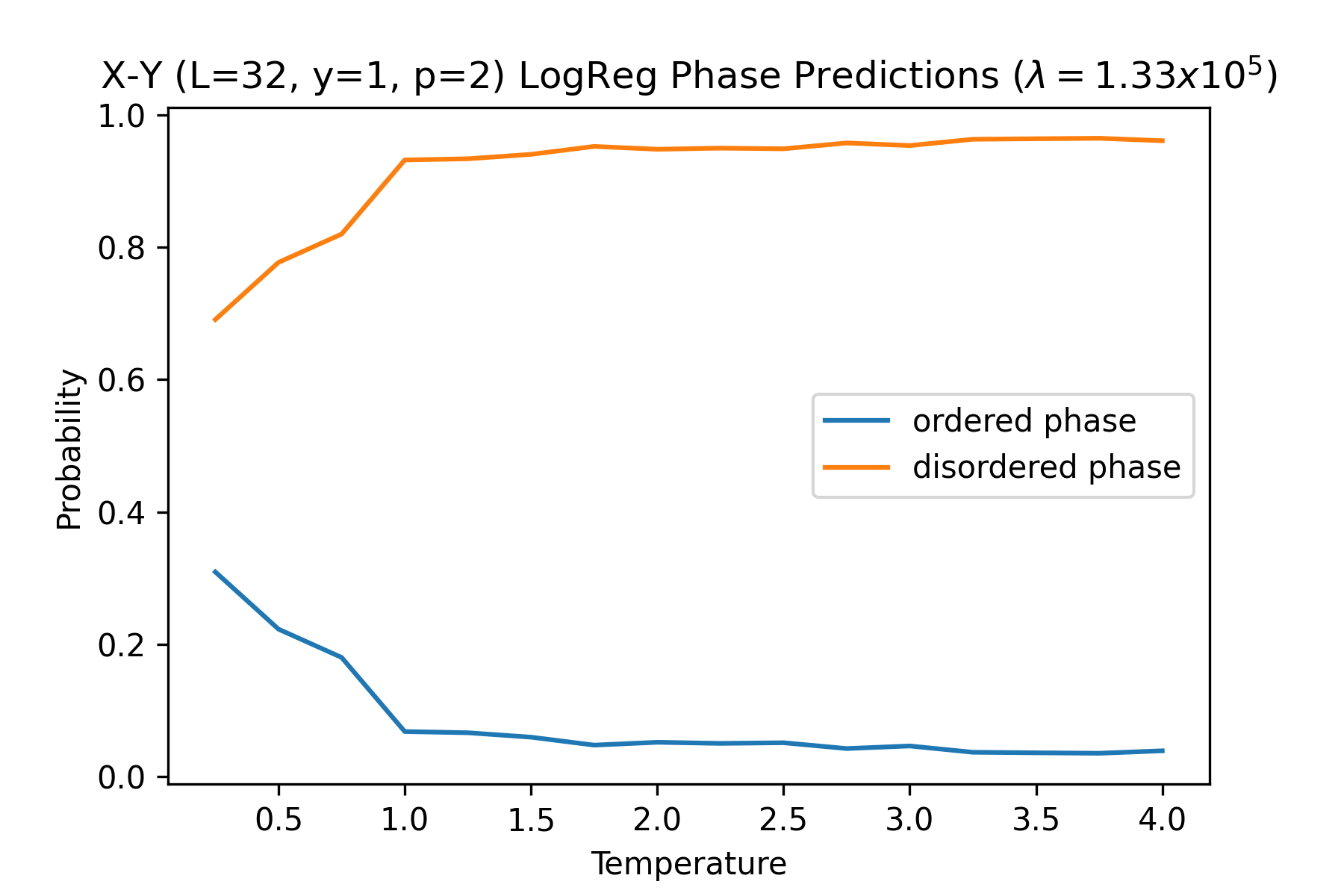}
\includegraphics[width=87mm]{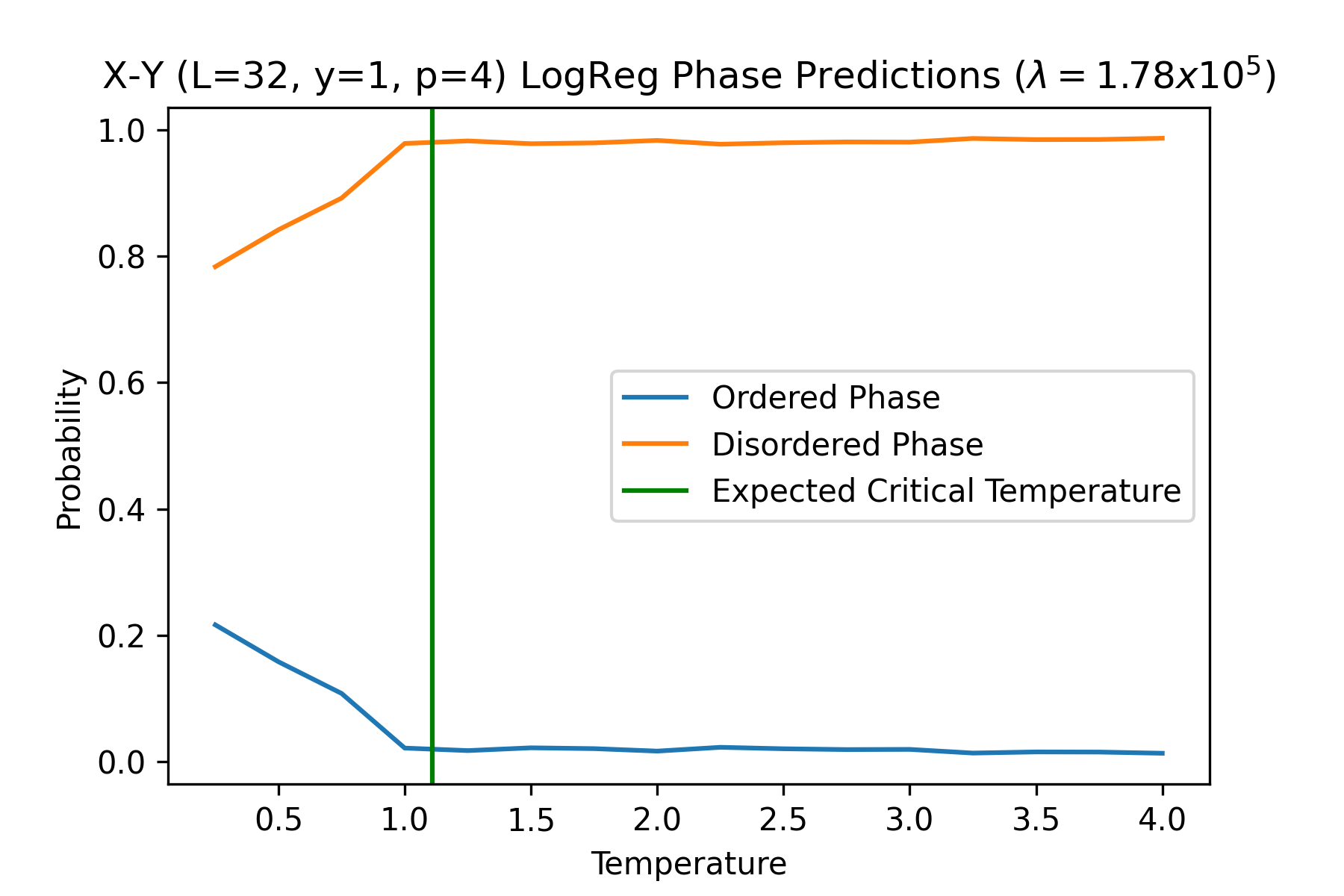}}
\caption{The prediction of the logistic regression algorithm applied to the pure XY-spin model with $\mathbb Z_2$- (top panel) and $\mathbb Z_4$- (bottom panel) preserving perturbations. We show the probabilities that the system is either in the ordered or disordered phase as a function of the dimensionless temperature $T$. We take $\lambda\cong 10^5$, which yields the maximum accuracy of $\cong 60\%$.  For example, we see from the right panel that the prediction of the logistic regression method is extremely poor.}
\label{prediction p 2 4}
\end{figure*}

We also use the same classifier to calculate the accuracy for the XY-spin model with $\mathbb Z_{p=2,4}$-preserving perturbations. The results are shown in FIG. \ref{accuracy p 2 4}. Here, unlike the pure XY-spin model, we find that a very poor accuracy is attained across a range of $\lambda$ that spans many orders of magnitude. Indeed, we completely fail to use the algorithm to see any transition, a result that is depicted in FIG. \ref{prediction p 2 4}. Similar poor results of the logistic regression model were reported in the literature when applied to the $\mathbb Z_2$ Ising model;  again, the model fails to predict the transition temperature accurately. It is interesting to see that this method fails whenever the model exhibits a discrete symmetry.

\subsection{Convolutional Neural Networks (CNN)}

A neural network is a specific machine learning model that attempts to mimic how a brain works. The network is composed of a series of nodes arranged into layers connected by synapses, with each having a specific weight that determines
how strongly it affects the output of the network. Each node in the network has an
activation function, which remaps the node's inputs non-linearly. The network
is trained using the stochastic gradient descent (SGD) method, which tries to
minimize a loss function that describes the error in the network’s output by adjusting
the weights of the synapses.
Convolutional neural networks (CNNs) are a special type of neural network commonly used to solve problems where spatial relationships within the data
are important, such as classifying images into categories. They are translationally invariant networks that respect the locality of the input data.   Due to the
interactions between the nearest lattice sites in the XY-spin models, it makes intuitive sense
that a CNN may be more effective at classifying lattices into ordered and disordered
phases. A CNN can be seen as taking a $2$-dimensional input and then applying a series
of filters to this input to extract features. The presence (or lack) of these features can
then be used to classify the input into a specific category.

There are two basic layers in a CNN: a convolution layer that computes the convolution of the input with a stack of filters and pooling layers that coarse-grain the input while maintaining locality. We use a CNN composed of a single convolutional layer with five $2\times2$ filters followed by a max pooling layer of kernel size $2\times2$ and a stride of $2\times 2$ yields near-perfect training and validation accuracy for all configurations of the XY-spin models.

\begin{figure*}[ht]
\leftline{
\includegraphics[width=87mm]{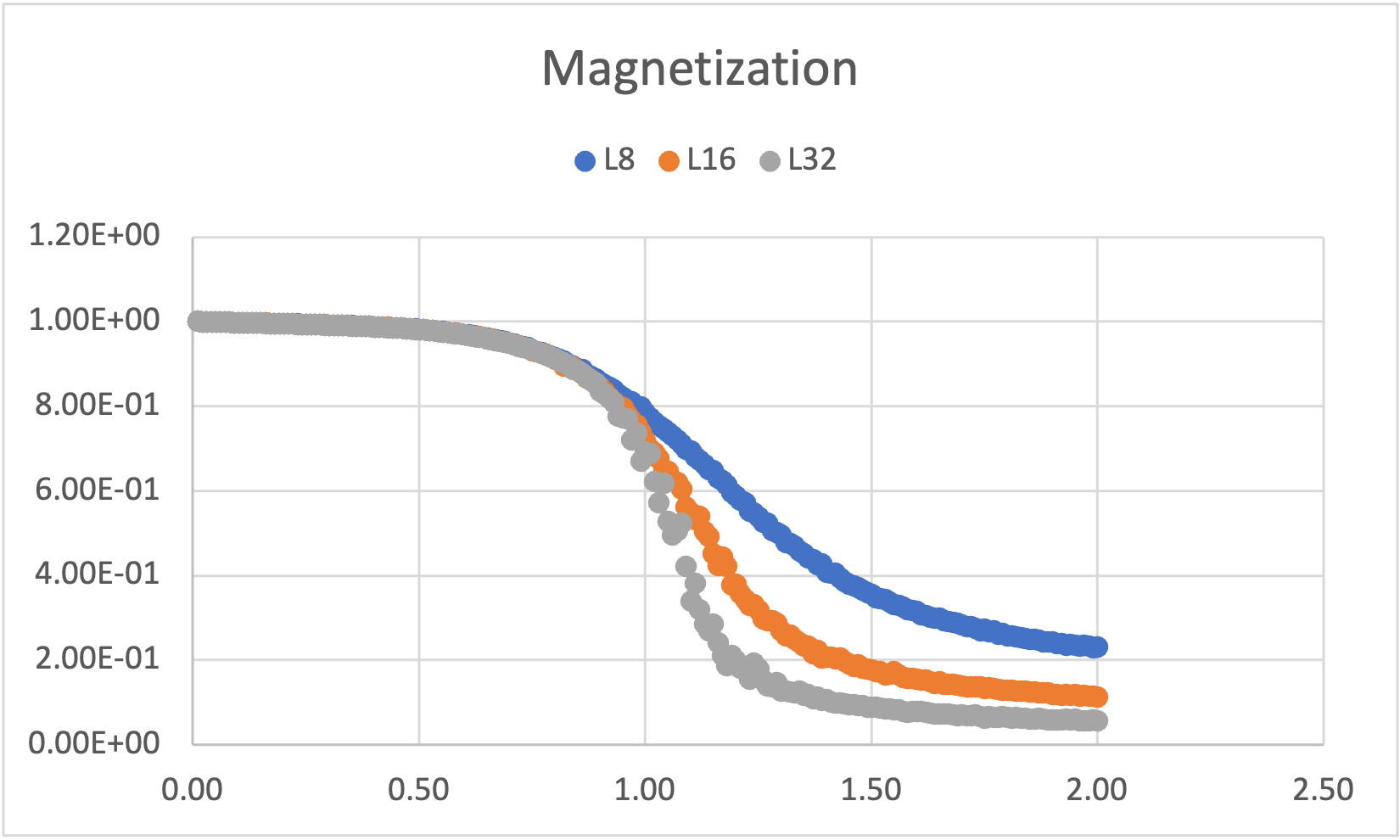}
\includegraphics[width=87mm]{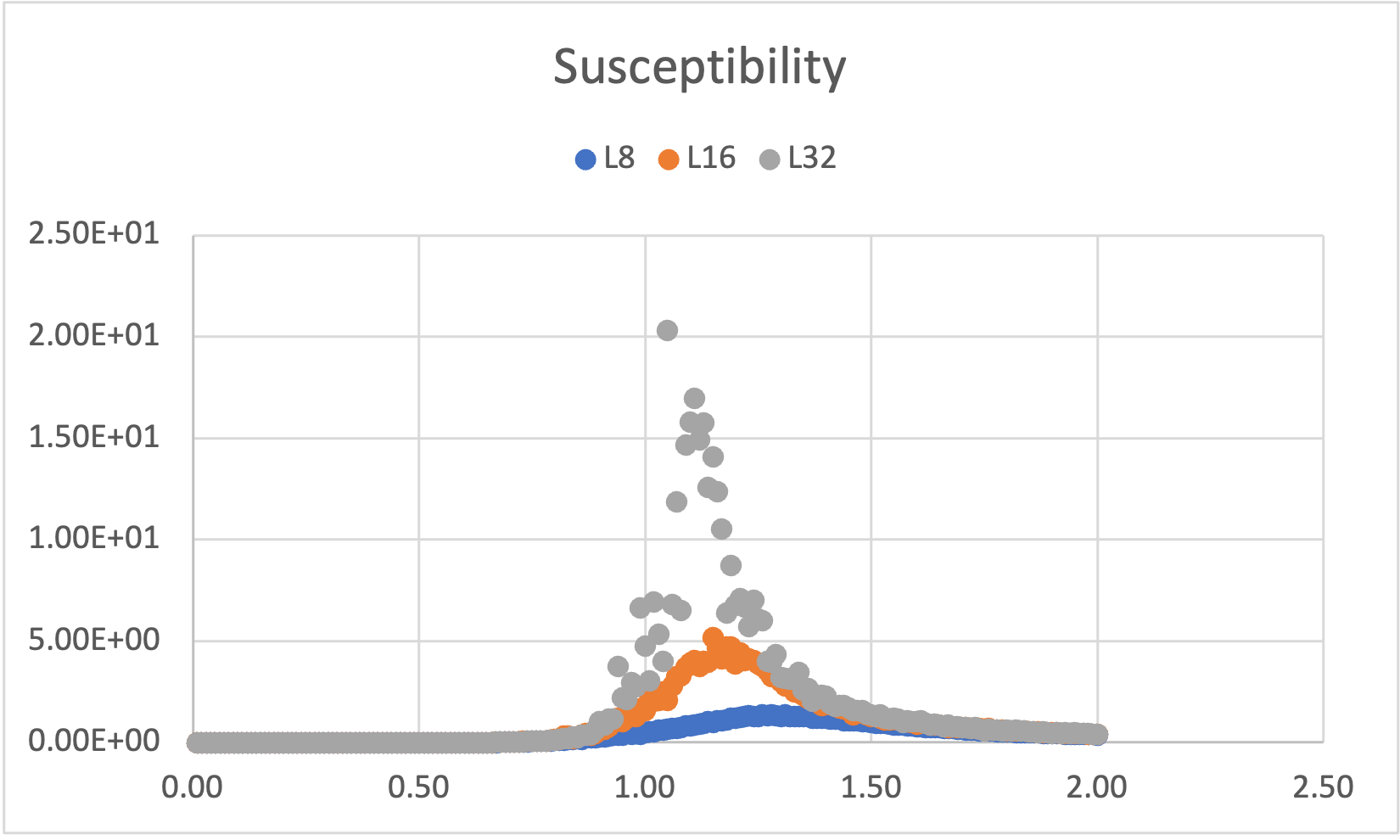}
}
\caption{The magnetization (left panel) and magnetic susceptibility (right panel) versus the dimensionless temperature $T$ of the XY-spin model with $y=1$ and $\mathbb Z_4$-preserving perturbations for lattice sizes $N=8,16,32$.}
\label{MS CNN 4}
\end{figure*}

To judge the accuracy of the CNN in predicting the transition temperature, we first review the results of the magnetic susceptibility for the XY-spin model with $y=1$ and $p=2,4$, which can be used to predict the transition temperatures using the finite-size scaling technique. The magnetization and magnetic susceptibility of a system are given by
\begin{eqnarray}
|M|\equiv\left\langle |\sum_{J=1}^{N^2}e^{i\theta_J}|\right\rangle\,, \quad \chi_M\equiv\frac{d|M|}{dT}\,.
\end{eqnarray}
Simulations of the XY-spin modules with $\mathbb Z_{2}$ and $\mathbb Z_4$ were performed in \cite{Anber:2018ohz} by one of the authors by varying the lattice sizes between $N=8$ and $N=56$.   The simulations indicated critical temperatures $T_c \cong1.53, 1.0$ for $y=1$ and $p=2,4$, respectively. As a reminder to the reader, $\mathbb Z_2$ and $\mathbb Z_4$ correspond to pure YM theory and YM theory with adjoint fermions, respectively. The state-of-art simulations for the XY-spin model with $\mathbb Z_4$-preserving perturbations at $y=1$ were performed in \cite{PhysRevB.69.174407} and yielded $T_c\cong 1.008+\pm0.002$. In FIG. \ref{MS CNN 4}, we plot the magnetization and the magnetic susceptibility for the $p=4, y=1$ case taking $N=8,16, 32$ lattice sizes. Although the accuracy of our simulations is relatively low compared to those existing in the literature, the peak of the magnetic susceptibility is at $ T\cong 1$, still consistent with more accurate simulations. Arguably, we show these data as a gauge against the predictions of the CNN, which are trained on the same Monte Carlo data used to produce FIG. \ref{MS CNN 4}.

\begin{figure*}[ht]
\leftline{
\includegraphics[width=87mm]{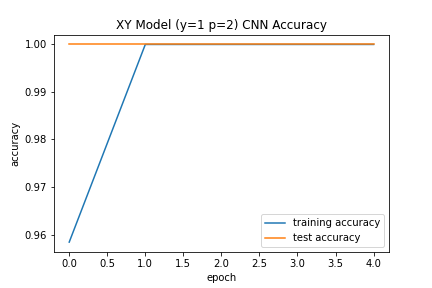}
\includegraphics[width=87mm]{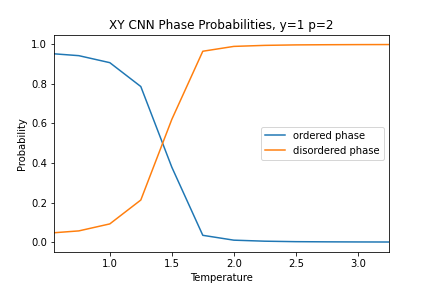}}
\caption{The accuracy (right panel) and prediction (left panel) of the CNN applied to the XY-spin model with $\mathbb Z_2$-preserving perturbations. We take $y=1$ and perform our simulations on a lattice size $N=40$.}
\label{CNN 2}
\end{figure*}

\begin{figure*}[ht]
\leftline{
\includegraphics[width=87mm]{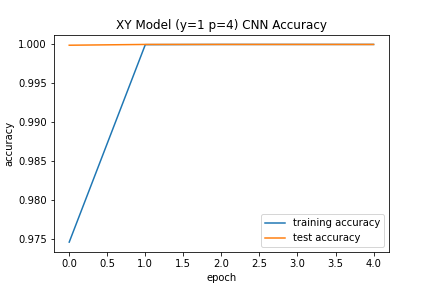}
\includegraphics[width=87mm]{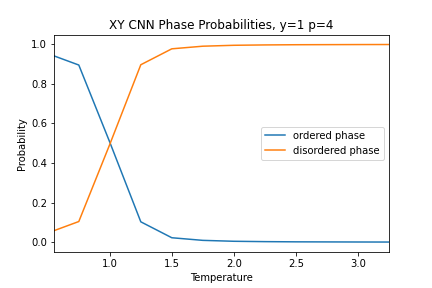}
}
\caption{The accuracy (right panel) and prediction (left panel) of the CNN applied to the XY-spin model with $\mathbb Z_4$-preserving perturbations. We take $y=1$ and perform our simulations on a lattice size $N=40$.}
\label{CNN 4}
\end{figure*}

Next, we discuss the results we obtained from training the CNN. We use a lattice of size $40\times 40$ and, as before, we take the temperature $T$ to range from $0.25$ to $4.0$ with a step size of $0.25$, for a total of $16$  temperatures and  $160,000$ configurations. The low-$T$  region is taken in the interval $[0.25, 0.75]$, the high-$T$ region is  $[3.25, 4.00]$, while the critical region is taken in the interval $(0.75, 3.25)$.  On the left panels of FIGs. \ref{CNN 2} and \ref{CNN 4}, we plot the accuracy of the CNN, for both the test and training data, for the XY-spin model with $y=1$ and $\mathbb Z_2$ and $\mathbb Z_4$, respectively. We plot the accuracy as a function of the number of epochs, where an epoch is a full iteration over the minibatches (collection of data points) used in the stochastic gradient descent minimization technique. Perfect accuracy is attained after training the CNN on the training data, and almost a $100\%$ accuracy is achieved on the test data. On the right panels of FIGs. \ref{CNN 2} and \ref{CNN 4}, we plot the probabilities that the system is either in the ordered or disordered phase. A $50\%$ chance that the system is in either phase predicts the transition temperature.  The predicted critical temperatures are $T_c \cong1.4$ and $T_c \cong1$ for $\mathbb Z_2$ and $\mathbb Z_4$, respectively, which agree with the critical temperatures found from the traditional methods.

%
\begin{table}
\center
\begin{tabular}{|c|c|c|c|}
\hline
$N$ &  $8$ &   $16$ &   $32$  \\
\hline\hline
 $\mbox{Max}\log \chi_M$ & $0.15$ &  $0.72$ &  $1.31$ \\
 \hline
 \hline
 $\mbox{Max}\log \chi_f$ & $0.89$ &  $1.51$ &  $2.21$ \\
\hline
\end{tabular}
\caption{The logarithm of the maxima of the magnetic and predictive function susceptibilities versus the logarithm of the lattice sizes $N=8,16,32$.}
\label{tab:max}
\end{table}
%

\begin{figure*}[ht]
\leftline{
\includegraphics[width=87mm]{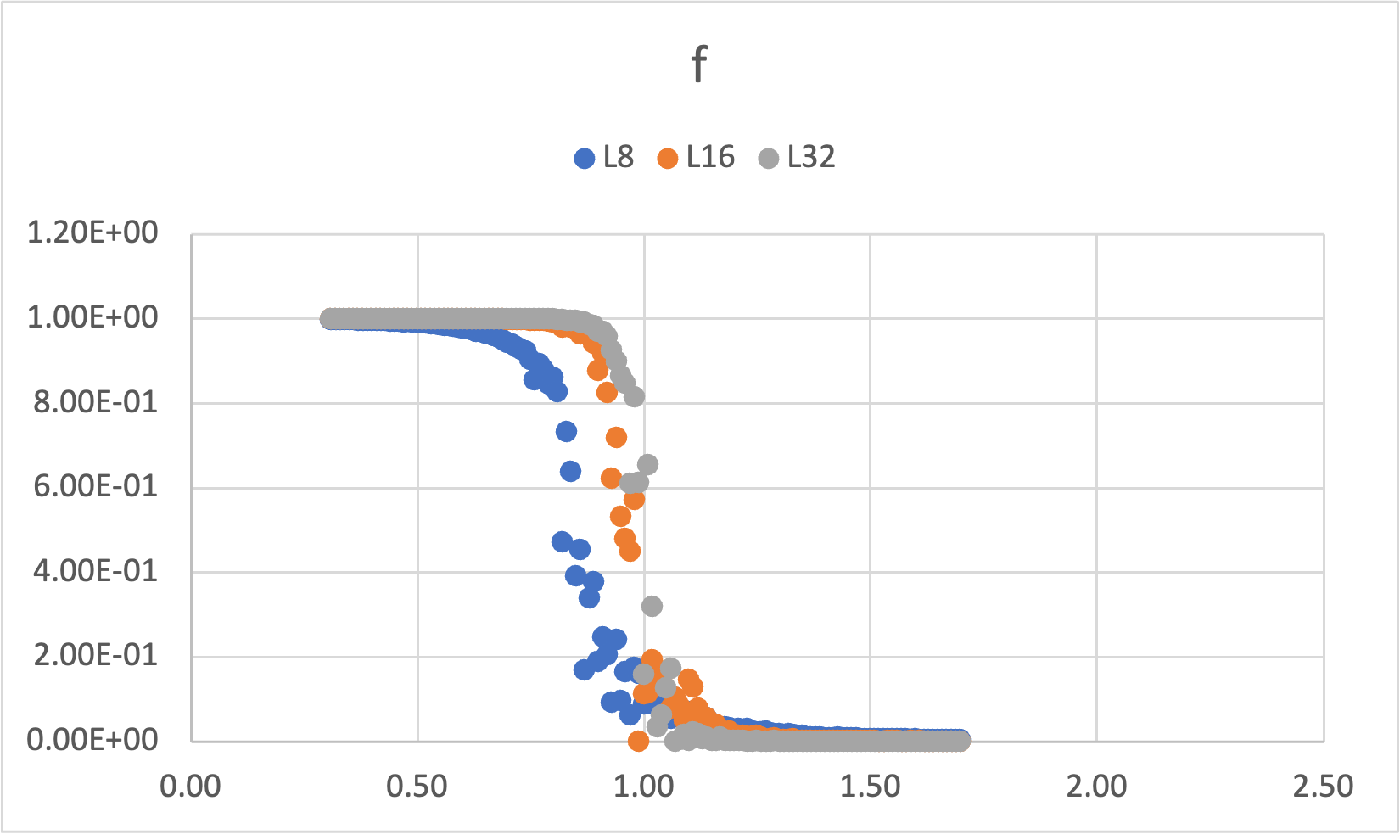}
\includegraphics[width=87mm]{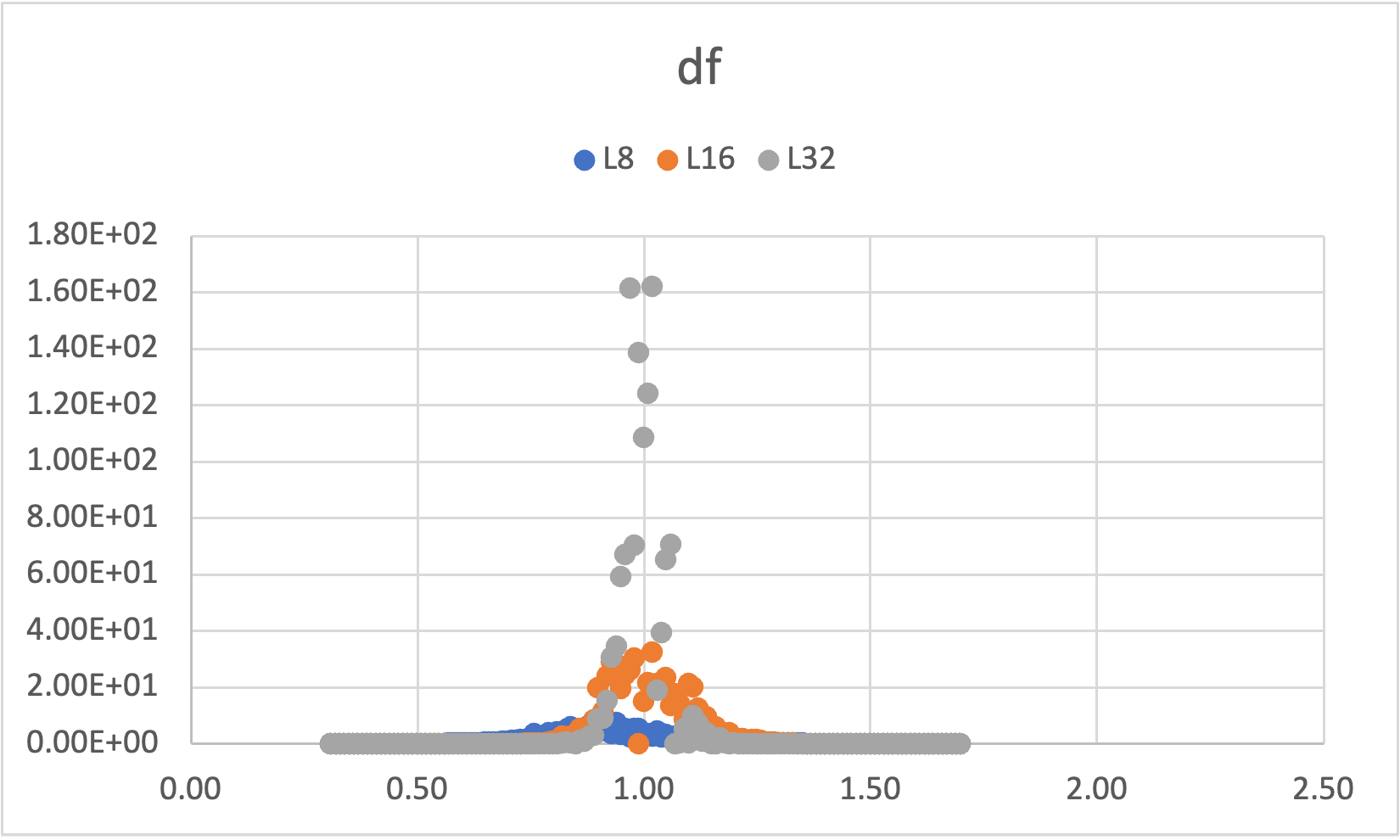}
}
\caption{The predictive function $f$ (left panel) and its derivative (right panel) versus the dimensionless temperature $T$ of the XY-spin model with $\mathbb Z_4$-preserving perturbations for lattice sizes $N=8,16,32$. We take $y=1$.}
\label{PF CNN 4}
\end{figure*}

\begin{figure*}[ht]
\leftline{
\includegraphics[width=87mm]{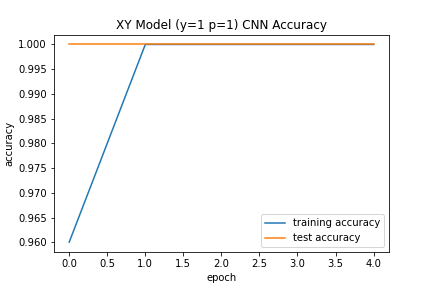}
\includegraphics[width=87mm]{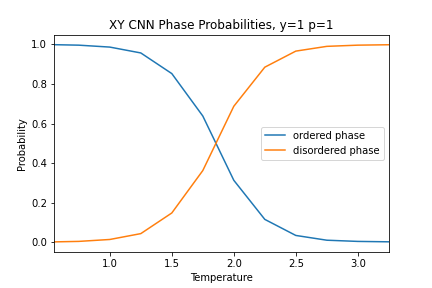}
}
\caption{The accuracy (right panel) and prediction (left panel) of the CNN applied to the XY-spin model with $y=1$ and $p=1$, which maps to YM theory with the fundamental matter. We perform our simulations on a lattice size $N=40$.}
\label{CNN 1}
\end{figure*}

%
\begin{table*}
\center
\begin{tabular}{|c|c|c|c|c|c|c|}
\hline
& $[0.1, 0.7]\cup[3.3,  4.0]$ &  $[0.1, 0.5]\cup[3.3, 4.0]$ &   $[0.1, 0.7]\cup[3.5, 4.0]$ &   $[0.1, 0.5]\cup[3.5, 4.0]$ & SD  \\
\hline\hline
$p=1, N=8$ & $1.65$ &  $1.47$ &  $1.86$ & $1.51$  &$0.18$ \\
  \hline
 $p=2, N=8$ & $1.341$ &  $1.336$ &  $1.343$ & $1.336$ & $0.004$ \\
 \hline
  $p=4, N=8$ & $0.849$ &  $0.847$ &  $0.851$ & $0.849$ & $0.002$ \\
  \hline\hline\hline
   $p=1, N=16$ & $1.76$ &  $1.70$ &  $1.83$ & $1.79$  &$0.05$ \\
   \hline
 $p=2, N=16$ & $1.456$ &  $1.451$ &  $1.454$ & $1.453$ & $0.002$ \\
 \hline
   $p=4, N=16$ & $0.971$ &  $0.961$ &  $0.973$ & $0.963$ & $0.006$ \\
  \hline\hline\hline
   $p=1, N=32$ & $1.87$ &  $1.75$ &  $1.90$ & $1.78$  &$0.07$ \\
   \hline
 $p=2, N=32$ & $1.543$ &  $1.54$ &  $1.545$ & $1.541$ & $0.002$ \\
  \hline
  $p=4, N=32$ & $0.959$ &  $0.957$ &  $0.959$ & $0.957$ & $0.001$ \\
   \hline\hline\hline
  $p=1, N=64$ & $1.48$ &  $1.24$ &  $1.87$ & $1.75$  &$0.28$ \\
 \hline
$p=2, N=64$ & $1.50$ &  $1.53$ &  $1.51$ & $1.46$ & $0.028$ \\
 \hline
 $p=4, N=64$ & $1.022$ &  $1.02$ &  $1.032$ & $1.01$ & $0.008$ \\
\hline
\end{tabular}
\caption{The transition temperatures and their standard deviation (SD) for the XY-spin models for $p=1,2,4$ on $N=8,16,32,64$ lattices, setting $y=1$, as we vary the training windows.}
\label{tab:training windows}
\end{table*}

To further quantify the predictions of the CNN versus the traditional method, we use the CNN to calculate the ratio of critical exponents $\gamma/\nu$ of the XY-spin model with $\mathbb Z_4$-preserving perturbations. Let $f(T)$ be the predictive function of a CNN at temperature $T$, which measures the probability that the system is in the ordered (low-temperature) phase. We define the susceptibility $\chi_f$ of the predictive function via the derivative:
\begin{eqnarray}
\chi_f\equiv \frac{df(T)}{dT}\,.
\end{eqnarray}
The magnetic and predictive function susceptibilities behave with the system size $N$ as $\chi\sim N^{\gamma/\nu}$; see, e.g.,  \cite{Bachtis:2020ajb}. 
Table \ref{tab:max} displays the logarithm of the maxima of the magnetic and predictive function susceptibilities versus the logarithm of the lattice sizes $N=8,16,32$. Using a least-square fit with a straight line, we obtain the following values of the ratio $\gamma/\nu$:
\begin{eqnarray}
\frac{\gamma}{\nu}=\left\{\begin{array}{cc} 1.93 & \quad\mbox{using}\, \chi_M \\ 2.2 &\quad\mbox{using}\,\chi_f\,.\end{array}  \right.
\end{eqnarray}
We find that the CNN prediction of $\gamma/\nu$ is only $14\%$ off the value obtained from the magnetization. As a reference point, we also mention that the state-of-art value of $\gamma/\nu=1.76\pm0.009$ was obtained in  \cite{PhysRevB.69.174407}. It is remarkable the CNNs give values of the critical exponents that are consistent with the exponents obtained from the physical order parameters. The fact that CNNs are never trained on data inside the critical region, and yet they can provide a good estimate of the thermodynamic properties of the systems is quite surprising. It will be interesting to check in future simulations whether one can achieve high accuracy in computing the critical exponents using ML techniques. 

Let us try to harness the CNN further, checking if they can provide a robust distinction of the phases of YM theory with fundamentals. It is well known that this theory does not possess order parameters, thanks to the fundamental fermions. Simply, no topological Wilson lines can be defined in this theory as they can end on the fundamental charges, hence the absence of a well-defined order parameter. As mentioned above, the XY-spin model with $p=1$ maps to YM theory with fundamentals. Thus, we repeat the above analysis, setting $y=1$ and $p=1$ in the model. As before,  the low-$T$  region is taken in the interval $[0.25, 0.75]$, the high-$T$ region is  $[3.25, 4.00]$, while the critical region is taken in the interval $(0.75, 3.25)$. We divide the data outside the critical region into training and test data and always ovoid training the CNN inside the critical region. The results are depicted in FIG. \ref{CNN 1}. As can be seen, the CNN attains high accuracy on the training and test data and predicts a transition temperature $T_c\cong 1.7$. Yet, one needs to examine the robustness of such transition temperature. If the system exhibits a true phase transition, the critical temperature should not strongly depend on the training data. To examine such dependence, we repeat our analysis for $p=1,2,4$ (setting $y=1$) while changing the boundaries of the training data. In TABLE \ref{tab:training windows},  we show how $T_c$ changes by changing the training windows on a $N=8,16,32,64$ lattice size. We consider two low-T windows $[0.1, 0.7]$, $[0.1, 0.5]$ and two high-T windows $[3.3,  4.0]$, [3.5, 4.0]. Then, we form $4$ distinct intervals built out of the low-T and high-T windows. For each lattice size, we observe a significantly greater variation in the transition temperature for the case with $p=1$ compared to the variations seen in the transition temperatures of the $p=2$ and $p=4$ cases as we manipulate the training windows. For example, for $N=64$, the critical temperature varies between $1.46$ and $1.53$ in the XY-spin model with $p=2$ (dYM), while it varies between $1.01$ and $1.03$  in the XY-spin model with $p=4$ (QCD(adj)). This variation is stunningly low, less than $0.02$ in the latter case, but still is very weak in the $p=2$ case, being less than the $0.07$ in dimensionless temperature units (remember that the Monte Carlo data are generated in step size $\Delta T=0.25$). On the other hand, the variation in the critical temperature in the case $p=1$ (fundamental fermions) is $0.61$ in dimensionless units, much bigger than the $\Delta T=0.25$ step size. We conclude that the transition temperature in the YM with fundamentals is not physically significant.

\section{Outlook}
\label{Outlook}

In this paper, we have applied the supervised ML techniques in phase transitions in YM theories defined on a small circle and endowed with center-stabilizing potential, without and with matter in the adjoint and fundamental representations. We have not tried to simulate the original theories, but used a duality that maps the original theories to XY-spin models with $\mathbb Z_p$-preserving perturbations. Distinct values of $p$ map to different YM theories: $p=2$ corresponds to pure YM, $p=4$ corresponds to YM theory with adjoint matter, and $p=1$ corresponds to YM with fundamentals. This map is an exact duality between the original YM theories and the XY-spin models, found using reliable semi-classical and effective field theory techniques,  and not a mere modeling of the original theories. The simulations of the XY-spin models are much simpler than the YM theories, especially if one wants to examine the suitability of the ML techniques in studying phase transitions.

While the logistic regression method proved to be unreliable in detecting phase transition in theories with $p=2,4$, CNNs, on the other hand, are found to be robust in obtaining the critical temperatures and critical exponents. We have also tried to check whether CNN can provide new insight on systems without global symmetries, the $p=1$ case. We concluded that the critical behavior observed using CNN is questionable since the critical temperature depends strongly on the training data. Any critical behavior should be robust against changing the boundaries of the windows of the training data. 

Recently, there has been an avalanche in the use of ML techniques, and there are various venues to extend our studies. For example, one can apply the persistent homology method, used in \cite{Sale:2021xsq} to study variants of the XY-spin model, to study XY spin-models with $\mathbb Z_p$ perturbations.  Another venue is the transfer learning technique, the breakthrough that gave rise to the celebrated ChatGPT \cite{radford2018improving}. Here, the features learned in the one given model can be used to predict the structure of symmetry-breaking phase transitions in other models, irrespective of the universality class. This method was applied in $q$-state Potts models in \cite{Bachtis:2020ajb}, and then the learned features were used to calculate the critical exponents in scalar field theory. Applying this method to the XY-spin models with $\mathbb Z_p$-preserving perturbations can reveal the scope of applicability of this method to classify a wide range of phases universally. 

\acknowledgements
 We would like to thank Erich Poppitz for comments on the manuscript. This work was supported in part by STFC through grant ST/T000708/1 and in part by NSF grant PHY-2013827. The simulations were performed on the BLT cluster at Lewis \& Clark College, Portland, USA. 

\bibliographystyle{apsrev4-1}
\bibliography{references}

\end{document}